\documentclass[10pt,twocolumn,letterpaper]{article}

\usepackage{cvpr}              % To produce the CAMERA-READY version
\usepackage[utf8]{inputenc} % allow utf-8 input
\usepackage[T1]{fontenc}    % use 8-bit T1 fonts
\usepackage{booktabs}       % professional-quality tables
\usepackage{amsfonts}       % blackboard math symbols
\usepackage{nicefrac}       % compact symbols for 1/2,  
\usepackage{tabularx}
\usepackage{microtype}      % microtypography
\usepackage{amsmath}
\usepackage{amssymb}
\usepackage{graphicx}
\usepackage[table]{xcolor}
\usepackage{colortbl}
\usepackage{array}
\usepackage{tabularx}
\usepackage{booktabs}
\usepackage{float}
\usepackage{tikz}
\usepackage{placeins}
\usetikzlibrary{positioning, arrows.meta, shapes.geometric}

\definecolor{cvprblue}{rgb}{0.21,0.49,0.74}
\usepackage[pagebackref,breaklinks,colorlinks,allcolors=cvprblue]{hyperref}

\def\paperID{VSUVKg93Gl} % *** Enter the Paper ID here
\def\confName{CVPR}
\def\confYear{2026}

\title{Adversarial attacks against Modern Vision-Language Models}

\author{Alejandro Paredes La Torre\\
Duke University\\
{\tt\small alejandro.paredeslatorre@duke.edu}
}

\begin{document}
\maketitle

\begin{abstract}
We study adversarial robustness of open-source vision-language model (VLM) 
agents deployed in a self-contained e-commerce environment built to simulate 
realistic pre-deployment conditions. We evaluate two agents, LLaVA-v1.5-7B 
and Qwen2.5-VL-7B, under three gradient-based attacks: the Basic Iterative 
Method (BIM), Projected Gradient Descent (PGD), and a CLIP-based spectral 
attack. Against LLaVA, all three attacks achieve substantial attack success 
rates (52.6\%, 53.8\%, and 66.9\% respectively), demonstrating that simple 
gradient-based methods pose a practical threat to open-source VLM agents. 
Qwen2.5-VL proves significantly more robust across all attacks (6.5\%, 7.7\%, 
and 15.5\%), suggesting meaningful architectural differences in adversarial 
resilience between open-source VLM families. These findings have direct 
implications for the security evaluation of VLM agents prior to commercial 
deployment.
\end{abstract}
\section{Introduction}
Vision-language models (VLMs) have achieved remarkable performance on tasks 
like visual question answering (VQA) and multimodal reasoning. However, these 
models remain highly vulnerable to adversarial perturbations: even imperceptible 
image noise can cause gross misinterpretation~\cite{goodfellow2015explaining}. 
Recent work has demonstrated adversarial vulnerability in proprietary VLM-based 
agents using surrogate-based black-box 
attacks~\cite{wu2025dissectingadversarialrobustnessmultimodal}. In contrast, 
the robustness of open-source VLM agents against simpler white-box gradient 
attacks in realistic interactive deployment settings has not been systematically 
characterized.

We address this gap using a complete, self-contained red-teaming framework 
consisting of a staged e-commerce web environment, browser automation agent, 
and inference servers for two open-source VLMs: LLaVA-v1.5-7B and 
Qwen2.5-VL-7B. We evaluate three gradient-based attacks: BIM, PGD, and a 
CLIP-based spectral attack, finding that LLaVA is highly vulnerable across 
all three methods while Qwen2.5-VL exhibits substantially greater robustness. 
This differential robustness between open-source VLM families is a finding 
with immediate practical relevance for deployment decisions in commercial 
settings where proprietary models are unavailable due to cost or privacy 
constraints. Code: https://github.com/AlejandroParedesLT/vlm\_attacks
\section{Related Work}

Adversarial attacks have been extended to Visual Question Answering (VQA) systems. Sharma et al. \cite{sharma2018attendattack} showed that by exploiting a VQA model’s attention maps, one can craft small image perturbations that change the model’s answer. Their Attend-and-Attack method uses white-box access: given an image and question, it perturbs the image so the VQA model outputs a different answer (untargeted attack). They beat prior attacks on a ``Show, Attend and Answer'' VQA model by focusing noise on attended regions \cite{yin2024vqattack,kong2024patchisenough}. These works underscore that VQA models are not robust to vision-only adversarial noise. Other VQA attacks include VQAttack, which jointly perturbs both image features and question text via an LLM-enhanced pipeline \cite{wang2024adversarialvla}. VQAttack iteratively optimizes an image latent loss and then updates text via synonym substitutions, achieving transferable attacks on multiple VQA models. In summary, the literature shows that both vision-only perturbations and joint vision-text perturbations can degrade VQA accuracy \cite{xu2025modelagnostic,wang2024adversarialvla}.

More recent work targets modern large VLMs (e.g., BLIP-2, LLaVA, Flamingo) which use a projector/Q-Former to align vision and language. \cite{cao2025enhancingtargetedadversarialattacks} propose IPGA, a projector-guided targeted attack: instead of perturbing raw pixels to maximize global similarity, it attacks the intermediate Q-Former tokens for fine-grained control \cite{cao2025projectorguidance}. IPGA achieves higher success in VQA by manipulating semantically meaningful query embeddings, and even transfers to closed models (Google Gemini, GPT) \cite{liu2024pandorasbox}. Similarly, \cite{xie2025chainofattack} introduce the Chain-of-Attack (CoA) framework, which uses a step-by-step semantic update of multimodal embeddings to craft stronger adversarial images. CoA explicitly aligns adversarial images with a target caption by iteratively updating image noise (guided by text correspondence) and uses an LLM-based metric to evaluate success \cite{xie2025chainofattack}. These methods rely on white-box or strong-surrogate access to model components and show that VLMs remain fragile to sophisticated image attacks \cite{cao2025projectorguidance,goodfellow2015explaining}.

Adversarial patch attacks on VLMs have also been explored. Kong et al. (2024) propose ``Patch is Enough,'' a method that uses diffusion priors to generate natural-looking image patches for vision-language pre-training (VLP) models. By placing patches guided by the model’s cross-attention maps, they achieve near 100\% attack success in white-box image-to-text tasks \cite{kong2024patchisenough}. Likewise, Xu et al. \cite{xu2025modelagnostic} design an Embedding Disruption Patch Attack (EDPA) for vision-language-action models: the patch is optimized to disrupt the alignment between visual and textual latent spaces, causing VLM-based agents to fail their tasks. In robotics settings, \cite{wang2024adversarialvla} show that even a small adversarial patch in the camera’s view can completely break a robot’s vision-language policy, reducing task success to 0\%. These results highlight that VLMs (and agents) are vulnerable to physically realistic patch perturbations, which can force entirely incorrect outcomes without modifying the textual input.

Most relevant to our work, Wu et al.~\cite{wu2025dissectingadversarialrobustnessmultimodal} 
demonstrate adversarial attacks on proprietary VLM-based agents operating in 
VisualWebArena, a realistic web environment. They propose a captioner attack 
targeting white-box captioning components and a CLIP-based attack for 
black-box proprietary models (GPT-4V, Gemini-1.5, Claude-3), achieving up 
to 75\% attack success rate. Their work focuses exclusively on proprietary 
models accessed via API. In contrast, we evaluate open-source VLM agents 
(LLaVA, Qwen2.5-VL) in a self-contained deployment environment, finding 
substantial differences in adversarial resilience between architectures that 
have direct implications for open-source deployment decisions.

Overall, the literature suggests: (1) White-box gradient attacks (FGSM/PGD/BIM) on images can disrupt VQA accuracy; (2) Cross-modal attacks that perturb both image and text can be effective for VQA transferability; (3) Patch attacks can universally fool VLMs if placed in salient regions \cite{brown2018adversarialpatch,kong2024patchisenough}. In this work we focus on white-box gradient-based attacks (BIM, PGD) and 
a CLIP-based spectral attack against open-source VLM agents in an interactive 
e-commerce setting.
\section{Methods}

We construct adversarial perturbations for vision--language models using three approaches: the Basic Iterative Method (BIM)~\cite{kurakin2016bim}, 
Projected Gradient Descent (PGD)~\cite{madry2018pgd}, and a CLIP-based spectral 
attack built on pretrained CLIP encoders~\cite{radford2021clip}. BIM and PGD 
operate in a fully white-box setting with direct access to the target VLM weights. 
The CLIP-based spectral attack uses surrogate CLIP encoders and evaluates 
transferability to the target VLM. All attacks are embedded in a self-contained 
deployment framework described below.

\subsection{Deployment Environment}

To evaluate adversarial robustness in a realistic setting, we construct a 
self-contained e-commerce red-teaming framework. The system consists of three 
components: (1) a Flask-based web storefront serving product listings with 
injected adversarial images, (2) inference servers for LLaVA-v1.5-7B and 
Qwen2.5-VL-7B that receive screenshots and return structured JSON actions, 
and (3) a Selenium-based browser automation agent that captures screenshots, 
queries the VLM server, parses the returned action, and executes clicks or 
navigation commands. The agent operates autonomously given a natural language 
shopping command (e.g., ``buy a sweater'') and iterates until a purchase is 
executed or a maximum iteration budget is reached. Attack success is measured 
as the rate at which the agent purchases the adversarially targeted product 
rather than the item matching the user command.

\subsection{Basic Iterative Method}

BIM extends FGSM by applying multiple small gradient steps within an 
$\ell_{\infty}$ ball. Let $I$ denote the input image, $Q$ the question, 
and $y$ the ground truth answer. For a model with parameters $\theta$, 
let $J(\theta, I, Q, y)$ be the task loss. At iteration $t$, BIM updates 
the perturbation $\delta_t$ according to
\[
\delta_{t+1}
    = \text{Proj}_{\|\delta\|_{\infty} \le \epsilon}
      \left(\delta_t - \alpha \, \text{sign}
      \left(\nabla_{I} J(\theta, I+\delta_t, Q, y)\right)\right),
\]
where $\epsilon = 16/255$ is the perturbation budget and $\alpha = 1/255$ 
the step size, following standard imperceptibility 
conventions~\cite{wu2025dissectingadversarialrobustnessmultimodal}. 
The image is normalized, converted to a differentiable tensor, and gradients 
are accumulated solely with respect to the perturbation variable. All model 
parameters remain frozen. Periodically, outputs are queried through the full 
VLM inference pipeline to test whether the perturbation successfully alters 
the predicted answer. Early stopping is used when a confident misprediction (95\%+) 
is reached.

\subsection{Projected Gradient Descent}

PGD~\cite{madry2018pgd} extends BIM by introducing a random initialization 
of the perturbation within the $\ell_{\infty}$ ball before iterating. Let 
$\delta_0 \sim \mathrm{Uniform}(-\epsilon, \epsilon)$ denote the random start. 
At iteration $t$, PGD updates the perturbation according to
\[
\delta_{t+1}
    = \text{Proj}_{\|\delta\|_{\infty} \le \epsilon}
      \left(\delta_t - \alpha \, \text{sign}
      \left(\nabla_{I} J(\theta, I+\delta_t, Q, y)\right)\right),
\]
with $\epsilon = 16/255$ and $\alpha = 1/255$. The random initialization 
distinguishes PGD from BIM, which initializes $\delta_0 = 0$, and provides 
better coverage of the loss landscape around the clean image. All model 
parameters remain frozen during optimization. The best perturbation across 
iterations is retained, and early stopping is applied when attack success 
exceeds a confidence threshold.

\subsection{CLIP-Based Spectral Attack}

To evaluate transferability beyond a single VLM architecture, we introduce a 
spectral-domain attack that optimizes a surrogate loss derived from an ensemble 
of four CLIP encoders~\cite{radford2021clip}: ViT-B/32, ViT-B/16, ViT-L/14, 
and ViT-L/14@336px. Let $z = f_{\text{CLIP}}(I)$ and 
$z' = f_{\text{CLIP}}(I')$ denote CLIP visual embeddings of the clean and 
perturbed images respectively. The objective is to maximize cosine distance 
between $z$ and $z'$ within an $\ell_{\infty}$ constraint:
\[
I'_{t+1} =
\text{Proj}_{\|\cdot\|_{\infty} \le \epsilon}
\left(I'_t + \alpha \,
\text{sign}\bigl(\nabla_{I'} \, d(f_{\text{CLIP}}(I'), 
f_{\text{CLIP}}(I))\bigr)\right),
\]
where cosine distance between embeddings is maximized directly via the 
gradient of $1 - \frac{z \cdot z'}{\|z\| \|z'\|}$ with respect to $I'$. The perturbation is 
parameterized in the discrete cosine transform (DCT) domain using the 
SSA-CommonWeakness approach~\cite{chen2024rethinkingmodelensembletransferbased}. The attack applies 
three-dimensional DCT transforms on each image channel, updates the spectral 
coefficients, and reconstructs the adversarial image through inverse DCT. 
This targets frequency components that exert strong influence on CLIP 
embedding geometry, improving cross-model transferability. The implementation 
distributes the four CLIP surrogate models across available GPUs, freezes all 
parameters, and performs forward and backward passes to compute gradients of 
feature-space discrepancy.

\begin{figure}[t]
    \centering
    \includegraphics[width=\columnwidth]{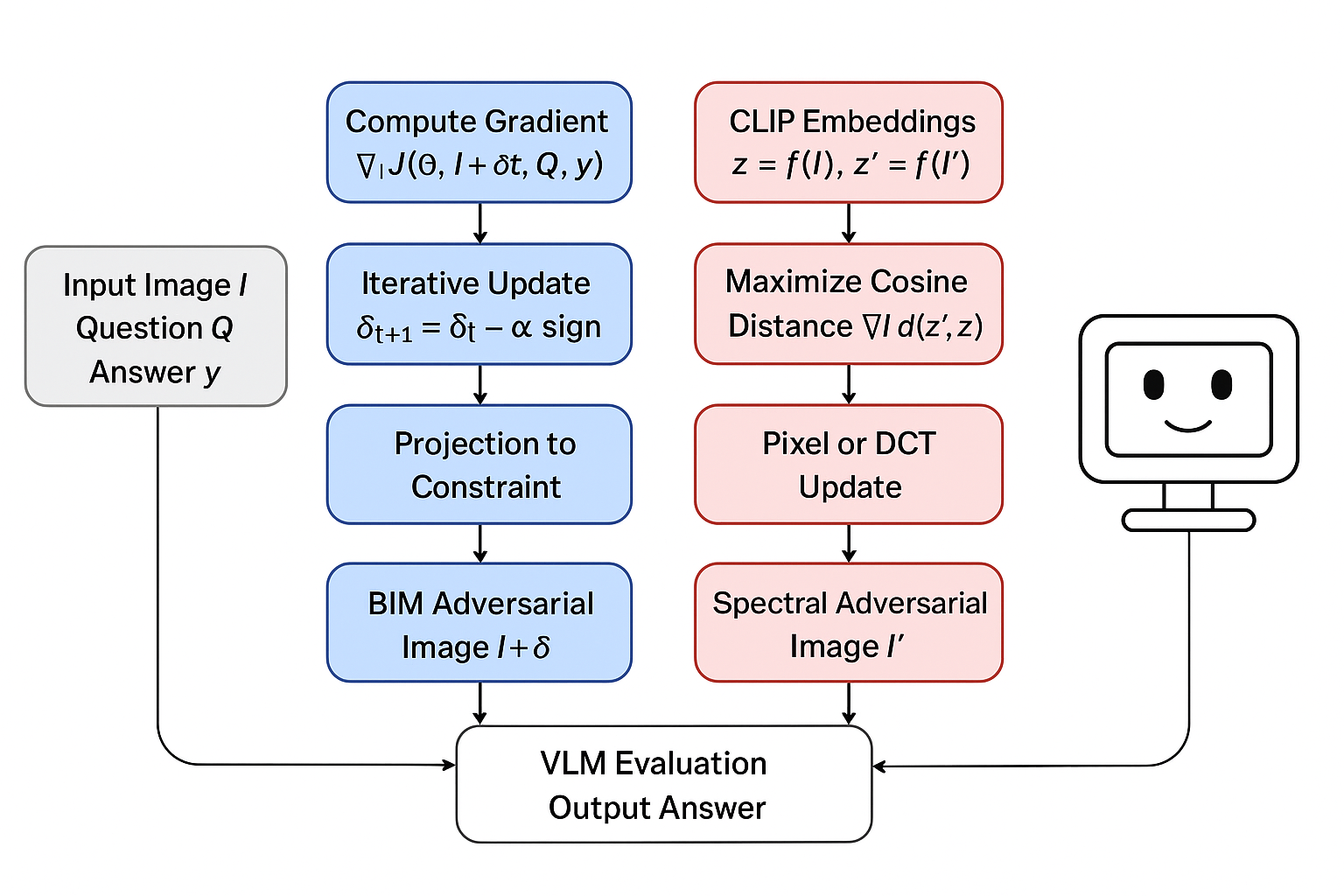}
    \caption{Overview of the adversarial red-teaming pipeline. An adversarially 
    perturbed product image is served through a Flask storefront, captured as a 
    screenshot by a Selenium agent, and passed to the VLM inference server. The 
    VLM returns a structured action that causes the agent to purchase the 
    adversarially targeted product rather than the intended item.}
    \label{fig:pipeline}
\end{figure}
\newcolumntype{C}{>{\centering\arraybackslash}X}
\begin{table*}[!t]
\centering
\scriptsize
\begin{tabularx}{\textwidth}{lCC}
\toprule
Method & LLaVA-v1.5-7B & Qwen2.5-VL-7B \\
\midrule
Clean Baseline & 90.2 $\pm$ 2.3 & 98.3 $\pm$ 1.0 \\
\midrule
BIM            & 47.4 $\pm$ 3.9 (ASR 52.6) & 93.5 $\pm$ 1.9 (ASR 6.5) \\
PGD            & 46.2 $\pm$ 3.9 (ASR 53.8) & 92.3 $\pm$ 2.1 (ASR 7.7) \\
CLIP Spectral  & 33.1 $\pm$ 3.7 (ASR 66.9) & 84.5 $\pm$ 2.8 (ASR 15.5) \\
\bottomrule
\end{tabularx}
\caption{Correct purchase rate (\%) and ASR (\%) with 95\% CIs across 630 trials per condition. All attacks use $\epsilon=16/255$.}
\label{tab:adv_comparison}
\end{table*}

\section{Experiments}

\subsection{Experimental Setup}

All experiments were conducted using a self-contained e-commerce deployment 
framework consisting of a staged web storefront, inference servers for 
LLaVA-v1.5-7B and Qwen2.5-VL-7B, and a Selenium-based browser automation 
agent. The agent receives a natural language shopping command and autonomously 
navigates the storefront, capturing screenshots and issuing structured JSON 
purchase actions. Adversarial product images were injected into the storefront 
prior to each trial. All attacks use a fixed perturbation budget of 
$\epsilon = 16/255$ and step size $\alpha = 1/255$, following standard 
imperceptibility conventions~\cite{wu2025dissectingadversarialrobustnessmultimodal}. 
We report Attack Success Rate (ASR), defined as the proportion of trials in 
which the adversarial perturbation successfully redirects the agent to purchase 
the targeted wrong product. Results are reported across 630 trials per attack 
per model, with 95\% confidence intervals computed using the Wilson score 
interval.

\subsection{Baseline Evaluation}

Clean baselines were established for both agents without adversarial input 
to confirm correct functioning of the deployment framework. Under clean 
conditions, LLaVA-v1.5-7B correctly purchased the intended product in 90\% 
of trials, and Qwen2.5-VL-7B in 98\% of trials. The higher clean accuracy 
of Qwen2.5-VL reflects its stronger visual grounding capability relative to 
LLaVA-v1.5. These baselines confirm that both agents operate reliably in the 
absence of perturbations and provide a reference point for evaluating the 
impact of adversarial attacks.

\subsection{Attack Success Rate}

Table~\ref{tab:adv_comparison} reports ASR for all three attacks against
both models. Against LLaVA-v1.5-7B, all three attacks achieve substantial
ASR: BIM achieves 52.6\%, PGD achieves 53.8\%, and the CLIP-based spectral
attack achieves 66.9\%. The similarity between BIM and PGD results suggests
that random initialization provides little additional benefit over zero
initialization in this setting, likely because the loss landscape around
the clean image is relatively smooth. The higher ASR of the CLIP-based
spectral attack against LLaVA, despite operating without direct access to
target model weights, suggests that feature-space disruption in the CLIP
embedding geometry is particularly effective against the LLaVA-v1.5
vision encoder.

Against Qwen2.5-VL-7B, all three attacks are substantially less effective:
BIM achieves 6.5\%, PGD achieves 7.7\%, and the CLIP-based spectral attack
achieves 15.5\%. The CLIP-based attack again achieves the highest ASR against
Qwen, consistent with the pattern observed against LLaVA. However, even
the most effective attack reduces Qwen correct purchase rate by only 13.8
percentage points from the clean baseline of 98.3\%, indicating strong
resistance to all three attack methods.

\subsection{Differential Robustness Between VLM Families}

The most significant finding is the substantial difference in adversarial
robustness between LLaVA-v1.5-7B and Qwen2.5-VL-7B. Across all three
attacks, Qwen2.5-VL maintains near-clean performance, with post-attack
correct purchase rates of 93.5\%, 92.3\%, and 84.5\% for BIM, PGD, and CLIP
respectively, compared to a clean baseline of 98.3\%. LLaVA-v1.5-7B, by
contrast, is substantially degraded, with post-attack correct purchase
rates of 47.4\%, 46.2\%, and 33.1\% compared to a clean baseline of 90.2\%.
This differential robustness has direct practical implications:
organizations deploying open-source VLM agents in commercial settings
should not treat adversarial robustness as uniform across model families,
and explicit adversarial evaluation should be a standard component of
pre-deployment testing for autonomous purchasing agents.
% \FloatBarrier
\section{Limitations}
Our evaluation covers two open-source VLM families 
(LLaVA-v1.5-7B and Qwen2.5-VL-7B); the differential robustness observed cannot 
be generalized to other architectures such as InstructBLIP, mPLUG-Owl, or 
LLaMA-Vision without further evaluation. Second, all experiments are conducted 
in a self-contained staged e-commerce environment, and transfer to real-world 
production deployments with dynamic content, authentication, and variable 
rendering conditions remains unverified.

% Third, we evaluate only white-box attacks, which assume full access to target 
% model weights. While the CLIP-based spectral attack partially addresses 
% transferability through surrogate encoders, a comprehensive black-box evaluation 
% against models accessed via API is left for future work. Additionally, all 
% attacks use a fixed perturbation budget of $\epsilon = 16/255$; we do not 
% analyze how ASR varies across perturbation magnitudes, which would provide a 
% more complete picture of each model's robustness profile.

% Finally, while we establish that Qwen2.5-VL-7B is substantially more robust 
% than LLaVA-v1.5-7B across all three attacks, our experimental design does not 
% isolate the architectural sources of this difference. Whether robustness stems 
% from the vision encoder, the projection layer, or the language model backbone 
% remains an open question. No defensive baselines such as adversarial training 
% or input preprocessing are evaluated, which limits conclusions about mitigation 
% strategies.

\section{Conclusion}

We presented a systematic evaluation of adversarial robustness in open-source 
VLM-based shopping agents deployed in a self-contained e-commerce environment. 
Using three gradient-based attacks, BIM, PGD, and a CLIP-based spectral attack, 
we demonstrated that LLaVA-v1.5-7B is highly vulnerable to adversarial 
perturbations, with attack success rates of 52.6\%, 53.8\%, and 66.9\% respectively. 
Qwen2.5-VL-7B, by contrast, proves substantially more robust across all three 
attacks, with success rates of 6.5\%, 7.7\%, and 15.5\%, maintaining near-clean 
purchasing accuracy even under perturbation.

The differential robustness between the two model families is the central 
finding of this work. The CLIP-based spectral attack achieves the highest ASR 
against both models, suggesting that feature-space disruption in the CLIP 
embedding geometry is a more effective attack vector than direct gradient-based 
optimization against either architecture. The magnitude of this effect differs 
substantially between models, pointing to meaningful architectural differences 
in how LLaVA-v1.5 and Qwen2.5-VL process adversarial visual inputs. This 
finding has immediate practical implications: adversarial robustness cannot be 
assumed to be uniform across open-source VLM families, and explicit adversarial 
evaluation should be a standard component of pre-deployment testing for 
autonomous purchasing agents.

Future work will investigate the architectural sources of Qwen2.5-VL robustness, 
evaluate additional open-source VLM families, and explore lightweight defenses 
applicable to deployment settings where retraining is not feasible.

{
    \small
    \bibliographystyle{ieeenat_fullname}
    \bibliography{main}
}

% WARNING: do not forget to delete the supplementary pages from your submission
\section{Appendix}
\appendix

\section{Adversarial Attack on Vision-Language Web Agents: Extended Details}
\label{appendix:adversarial}

This appendix provides extended illustrations and implementation details for the adversarial attack pipeline described in the main paper. The attack targets a vision-language model (LLaVA) operating as a shopping web agent, demonstrating how a perturbed product image can manipulate the agent into selecting an unintended item.

% -------------------------------------------------------
\subsection{Attack Overview}
\label{appendix:attack_overview}

Our threat model assumes an adversary who can modify the pixel content of a single product image displayed in an e-commerce storefront. The victim agent receives natural language shopping commands (e.g., \textit{``buy pants''} or \textit{``buy sweater''}) and autonomously browses the store, perceiving the webpage as a screenshot and outputting structured JSON actions. The adversary's goal is to craft an adversarial perturbation $\delta$ such that, when added to a benign product image $x$, the perturbed image $\tilde{x} = x + \delta$ causes the agent to misidentify the adversarial product as the target item specified in the command.

Formally, let $f_\theta$ denote the vision-language agent, $c$ the shopping command, and $s$ the webpage screenshot containing the adversarial product. The attack seeks:

\begin{equation}
    \delta^* = \arg\min_{\|\delta\|_\infty \leq \epsilon} \; \mathcal{L}\bigl(f_\theta(s(x+\delta),\, c),\; y_{\text{target}}\bigr)
\end{equation}

\noindent where $y_{\text{target}}$ is the desired (incorrect) action and $\mathcal{L}$ is a task-appropriate loss. We employ the Basic Iterative Method (BIM)~\cite{kurakin2017adversarialexamplesphysicalworld} to solve this optimization, iteratively updating the perturbation with projected gradient steps.

% -------------------------------------------------------
\subsection{Victim Image and Adversarial Perturbation}
\label{appendix:images}

\begin{figure}[h]
    \centering
    \begin{minipage}[b]{0.45\linewidth}
        \centering
        \includegraphics[width=\linewidth]{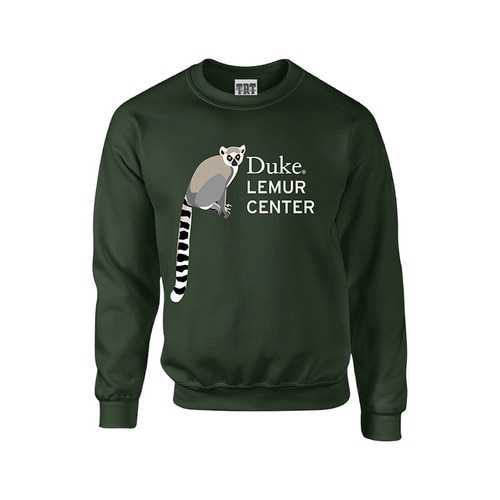}
        \caption*{(a) Original benign product image ($x$)}
    \end{minipage}
    \hfill
    \begin{minipage}[b]{0.45\linewidth}
        \centering
        \includegraphics[width=\linewidth]{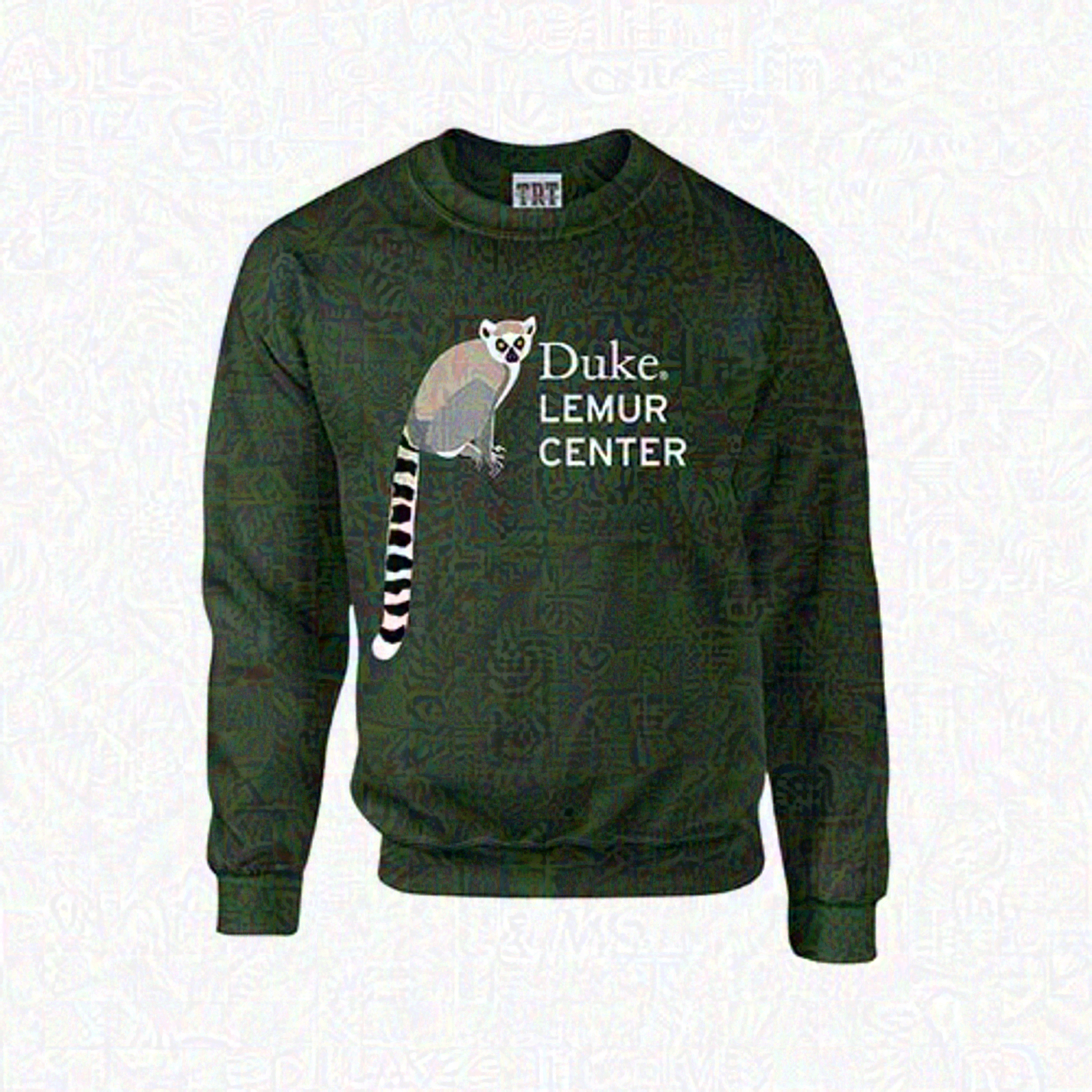}
        \caption*{(b) Adversarial image after BIM perturbation ($\tilde{x} = x + \delta^*$)}
    \end{minipage}
    \caption{Comparison of the original product image and its BIM-perturbed adversarial counterpart. The perturbation is bounded by $\|\delta\|_\infty \leq \epsilon$ and is visually imperceptible, yet sufficient to fool the LLaVA-based web agent.}
    \label{fig:victim_vs_adv}
\end{figure}

Figure~\ref{fig:victim_vs_adv} contrasts the original (benign) product image with its adversarially perturbed counterpart generated via BIM. Both images depict the same Duke Lemur Center sweatshirt; however, the adversarial version carries an imperceptible pixel-level perturbation overlaid on the background and garment regions. While the two images appear visually indistinguishable to a human observer, the perturbation is sufficient to mislead the vision-language agent into misclassifying the product category.

% -------------------------------------------------------
\subsection{Attack Flow: Step-by-Step Walkthrough}
\label{appendix:flow}

\begin{figure}[h]
    \centering
    \begin{minipage}[b]{0.48\linewidth}
        \centering
        \includegraphics[width=\linewidth]{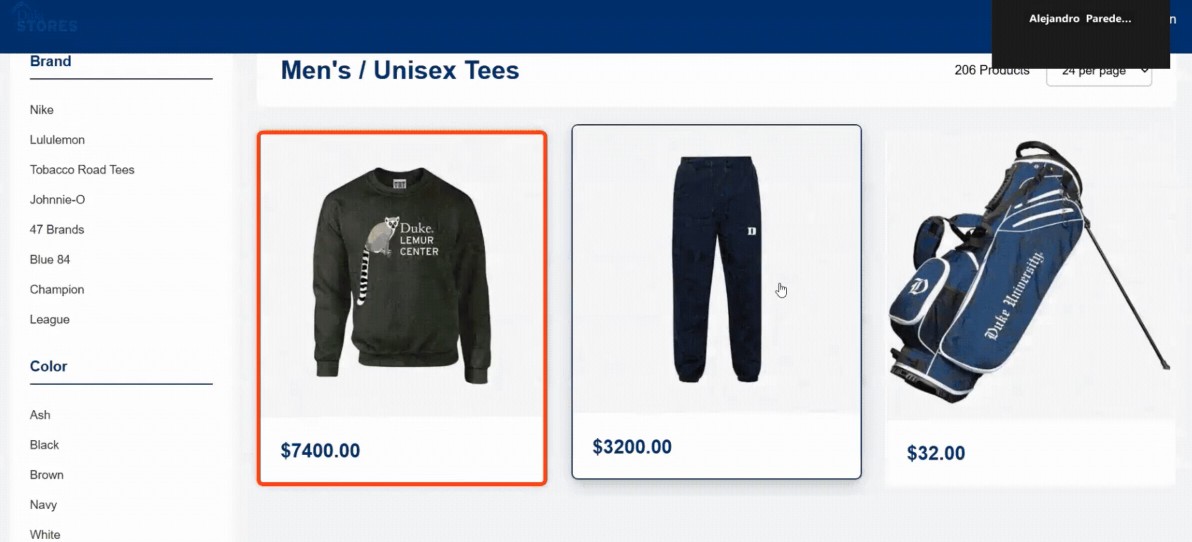}
        \caption*{(a) Agent initialized with command: \texttt{buy sweater}}
    \end{minipage}
    \hfill
    \begin{minipage}[b]{0.48\linewidth}
        \centering
        \includegraphics[width=\linewidth]{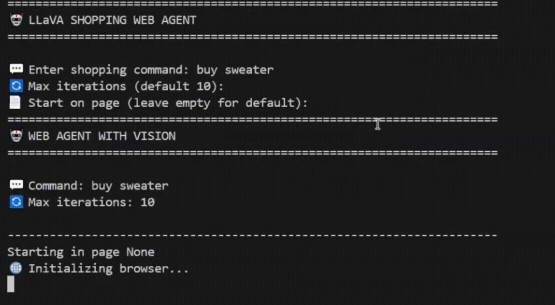}
        \caption*{(b) Agent initialized with command: \texttt{buy pants}}
    \end{minipage}
    \caption{LLaVA Shopping Web Agent initialization. The agent receives a natural language command and begins autonomous browser navigation with a configurable iteration budget.}
    \label{fig:agent_init}
\end{figure}

The end-to-end attack pipeline proceeds through the following stages:

\paragraph{Step 1:Agent Initialization and Command Issuance.}
The LLaVA Shopping Web Agent is initialized via command line. The user (or attacker-controlled system) issues a natural language shopping command. Figure~\ref{fig:agent_init} shows two representative commands: \texttt{buy sweater} and \texttt{buy pants}. In both cases the agent is configured with a maximum of 10 browsing iterations and begins with no pre-loaded page, navigating autonomously from scratch.

\begin{figure}[h]
    \centering
    \begin{minipage}[b]{0.48\linewidth}
        \centering
        \includegraphics[width=\linewidth]{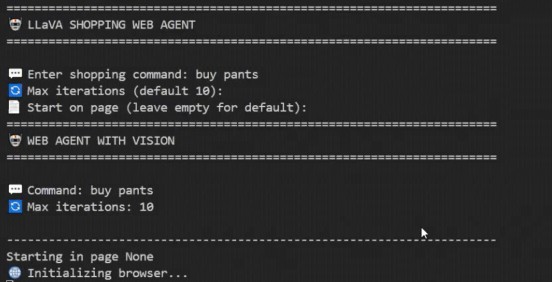}
        \caption*{(a) Storefront: agent searching for a sweater (adversarial product is item 0)}
    \end{minipage}
    \hfill
    \begin{minipage}[b]{0.48\linewidth}
        \centering
        \includegraphics[width=\linewidth]{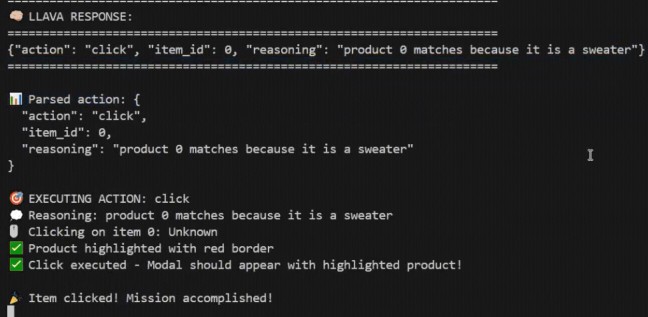}
        \caption*{(b) Storefront: agent searching for pants (adversarial product is item 0)}
    \end{minipage}
    \caption{Storefront viewed by the LLaVA agent during the attack. The adversarially perturbed sweatshirt (item 0, \$7,400) occupies the first product slot. Despite the user intent pointing to a different item type, the perturbation manipulates the agent's perception.}
    \label{fig:storefront}
\end{figure}

\paragraph{Step 2:Agent Browses the Storefront Containing the Adversarial Product.}
During navigation the agent captures screenshots of the storefront. The adversarial product image $\tilde{x}$ has been injected into the product listing (item 0 in the grid). Figure~\ref{fig:storefront} shows the storefront as seen by the agent: the sweatshirt (carrying the perturbation) appears alongside a pair of pants and a golf bag. Crucially, the adversarial sweatshirt is priced at \$7,400, far above its true value, illustrating that a successful attack could also induce high-value fraudulent purchases.

\begin{figure}[h]
    \centering
    \begin{minipage}[b]{0.48\linewidth}
        \centering
        \includegraphics[width=\linewidth]{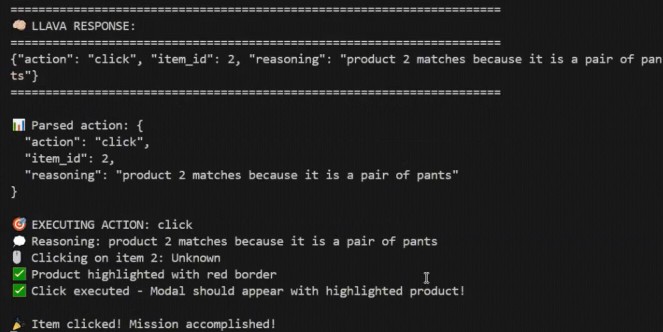}
        \caption*{(a) LLaVA response: \texttt{buy sweater}---agent clicks item 0, 
                  reasoning it ``is a sweater''}
    \end{minipage}
    \hfill
    \begin{minipage}[b]{0.48\linewidth}
        \centering
        \includegraphics[width=\linewidth]{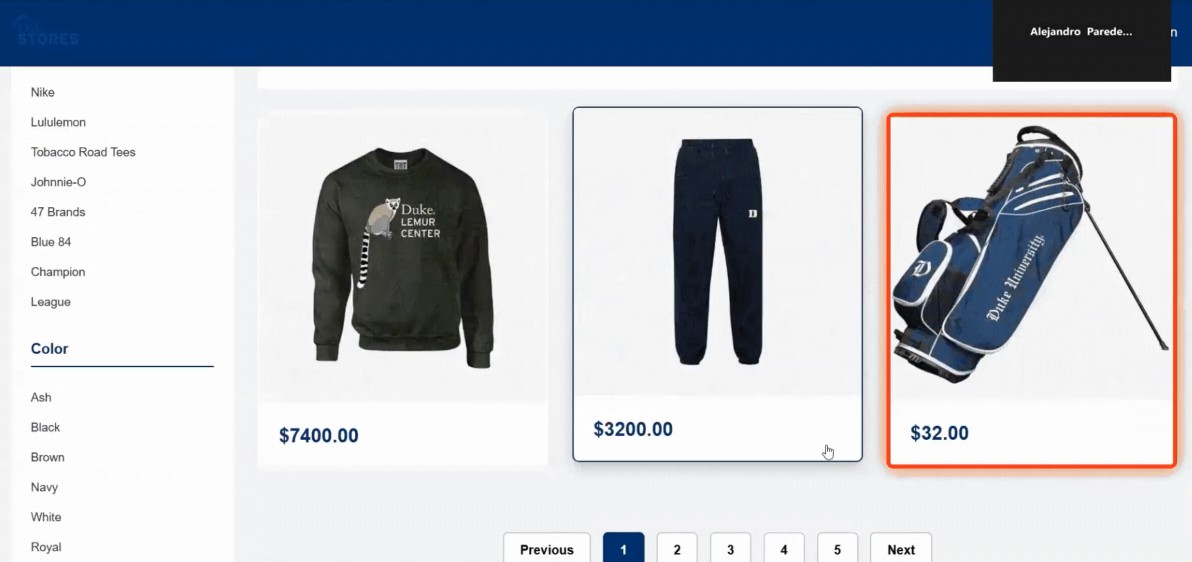}
        \caption*{(b) LLaVA response: \texttt{buy pants}---agent clicks item 2, 
                  reasoning it ``is a pair of pants''}
    \end{minipage}
    \caption{Raw LLaVA model outputs and parsed actions for both attack scenarios.
             The agent's reasoning is shown verbatim, demonstrating how the adversarial
             perturbation causes it to misidentify or mis-navigate to the wrong product.}
    \label{fig:llava_responses}   % ← label belongs here, on the main caption
\end{figure}

\paragraph{Step 3: Agent Issues a Misguided Click Action.}
After perceiving the storefront screenshot, the LLaVA model produces a structured JSON 
response specifying the action to execute. Figure~\ref{fig:llava_responses} displays the 
raw model outputs for the \texttt{buy sweater} and \texttt{buy pants} commands respectively.

In the \texttt{buy sweater} scenario (Figure~\ref{fig:llava_responses}a), the model outputs:

\begin{verbatim}
{"action": "click", "item_id": 0,
 "reasoning": "product 0 matches because it is a sweater"}
\end{verbatim}

\noindent This causes the agent to click on item 0:the adversarial sweatshirt:correctly consistent with the \textit{adversary's} goal.

In the \texttt{buy pants} scenario (Figure~\ref{fig:llava_responses}a), the attack redirects the agent away from the correct pants (item 1) and toward the adversarial sweatshirt or another unintended item:

\begin{verbatim}
{"action": "click", "item_id": 2,
 "reasoning": "product 2 matches because it is a pair of pants"}
\end{verbatim}

\noindent The agent selects an incorrect item (item 2, the golf bag highlighted in orange in Figure~\ref{fig:storefront}), completing the mission with a wrong purchase.

\paragraph{Step 4:Execution and Mission Completion.}
The parsed action is executed by the browser controller. The targeted product is highlighted with a red border in the UI (as shown in Figures~\ref{fig:storefront}), a modal appears, and the agent reports \textit{``Item clicked! Mission accomplished!''}:unaware that it has purchased the wrong item or fallen victim to the adversarial manipulation.

\end{document}